\documentclass{iaus}
\usepackage{graphicx}

\title[Transit observations in Gro{\ss}schwabhausen: TrES-2] 
{Observations of the transiting planet TrES-2 with the AIU Jena telescope in Gro{\ss}schwabhausen}

\author[St. Raetz et al.]   
{St. Raetz$^1$,
M. Mugrauer$^1$,
T. O. B. Schmidt$^1$,
T. Roell$^1$,
\\ T. Eisenbeiss$^1$,
M.M. Hohle$^{1,4}$,
A. Seifahrt$^{1,2}$,
A. Koeltzsch$^1$,
M. Va{\v n}ko$^1$,
Ch. Broeg$^{3}$,
J. Koppenhoefer$^5$
\and R. Neuh{\"a}user$^1$}

\affiliation{$^1$Astrophysikalisches Institut und Universit{\"a}ts-Sternwarte Jena, Schillerg{\"a}sschen 2-3, 07745 Jena, Germany; email: straetz@astro.uni-jena.de \\[\affilskip]
$^2$Institut f{\"u}r Astrophysik, Georg-August-Universit{\"a}t, Friedrich-Hund-Platz 1, 37077 G{\"o}ttingen, Germany \\[\affilskip]
$^3$Space Research and Planetary Sciences, Physikalisches Institut, University of Bern, Sidlerstraße 5, 3012 Bern, Switzerland \\[\affilskip]
$^4$Max-Planck-Institut of Extraterrestial Physics, Giessenbachstrasse, 85748 Garching, Germany\\[\affilskip]
$^5$University Observatory Munich, Scheinerstrasse 1, 81679 M\"unchen, Germany}

\pubyear{2008}
\volume{253}  
\pagerange{119--126}
\jname{Transiting Planets}
\editors{A.C. Editor, B.D. Editor \& C.E. Editor, eds.}
\begin{document}

\maketitle

\begin{abstract}
We have started high precision photometric monitoring observations at the AIU Jena observatory in Gro{\ss}schwabhausen near Jena in fall 2006. We used a 25\,cm Cassegrain telescope equipped with a CCD-camera mounted picky-pack on a 90\,cm telescope. To test the obtainable photometric precision, we observed stars with known transiting planets. We could recover all planetary transits observed by us.\\ We observed the parent star of the transiting planet TrES-2 over a longer period in Gro{\ss}schwabhausen. Between March and November 2007 seven different transits and almost a complete orbital period were analyzed. Overall, in 31 nights of observation 3423 exposures (in total 57.05\,h of observation) of the TrES-2 parent star were taken. Here, we present our methods and the resulting light curves. Using our observations we could improve the orbital parameters of the system.
\end{abstract}

\firstsection 
\section{Introduction}

The search for extrasolar planets is one of the most important research field in Astronomy today. Up to now the most successful method to detect exoplanet candidates is the radial velocity technique. But in the last ten years another indirect detection method has established itself as a highly successful technique in finding planets and confirming candidates - the transit method. \\ The transit event is a strictly periodic phenomenon. In a systen where a known planet transits its host star, a second planet in that system will cause the time between transits to vary. It is becoming increasingly popular because even small ground-based observatories have already obtained the photometric precision necessary to detect sub-Earth mass planets by the transit timing variation method \cite[(Steffen et al. 2007)]{Steffen_etal07}.\\ We have started high precision photometric monitoring observations at the AIU Jena observatory in Gro{\ss}schwabhausen near Jena in fall 2006. In this work we use the transit method to observe the known transiting planet TrES-2. The aim is to test procedures with our telescope and determine the obtainable photometric precision with the currently existing camera. We paid special attention to the accurate determination of transit times in order to identify precise transit timing variations that would be indicative of perturbations from additional bodies and to refine the orbital parameters of the systems.

\section{Instruments and Observations}

We have three telescopes available in our observatory in Gro{\ss}schwabhausen, a 90\,cm reflector, a 20\,cm refractor with a focal length of 3\,m and a 25\,cm Cassegrain telescope with a focal ratio f/D\,=\,9. The 90\,cm reflector telescope made by Zeiss Jena can be used in two modes -– either as Schmidt camera (diameter of the correction plate D\,=\,60\,cm, f/D\,=\,3) or as Nasmyth telescope with D\,=\,90\,cm of free opening and f/D\,=\,15.\\ Because new motors for the movement of the telescope were installed, we currently test procedures with the 25\,cm Cassegrain telescope with the CCD-camera \textit{CTK} (\textit{\underline{C}assegrain-\underline{T}elescop CCD-\underline{K}amera})(\cite[Mugrauer et al. (2009)]{Mugrauer_etal09}, in preperation). In the course of the year 2006 we started our continuous observations. \\ For our TrES-2 observations, started in March 2007, we used 34 nights from March to November 2007. Due to the weather conditions 3 nights were not analyzed. All TrES-2 observations were taken in I-band with 60 s exposure time. The mean photometric accuracy of the V\,=\,11.4\,mag bright TrES-2 host star is 0.007 mag.

\section{Data Reduction and analysis}

\begin{figure}[h]
  \centering
  \includegraphics[width=1\textwidth]{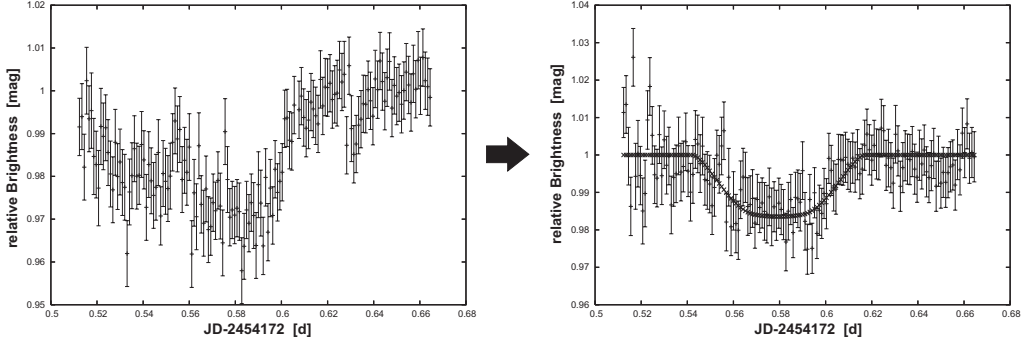}
  \caption{The same transit of TrES-2 observed on March 13 2007 before and after using Sys-Rem. The time series includes 148 I-band 60s exposures between 0.17 AM and 3.56 AM (UT)}
  \label{TrES_Sysrem}
\end{figure}

We calibrate the images of our target fields using the standard IRAF\footnote{IRAF is distributed by the National Optical Astronomy Observatories, which are operated by the Association of Universities for Research in Astronomy, Inc., under cooperative agreement with the National Science Foundation.} procedures \textit{darkcombine}, \textit{flatcombine} and \textit{ccdproc}.\\ After calibrating all images, we perform aperture photometry. Therefore we use the IRAF task \textit{chphot} which is written by Christopher Broeg and based on the standard IRAF routine \textit{phot}. With \textit{chphot} it is possible to do the photometry on every star in the field at the same time. We found 1294 stars in our 37.7\,'$\times$ 37.7\,'field of view around TrES-2. We used an aperture of radius 5 pixels (11.03'') and an annulus for sky subtraction ranging in radius from 15 to 20 pixels, centered on each star.\\ A problem of the differential photometry is the search for a good comparison star. \cite[Broeg et al. (2005)]{Broeg_etal05} developed an algorithm which uses as many stars as possible (all available field stars) and calculate an artificial comparison star (cs). The algorithm decides which stars are the best and takes the weighted average of them. Then it computes the artificial cs with the best possible signal-to-noise ratio (S/N) by automatically weighting down the stars according to their variability.\\ To get the best possible result for the transit light curve, we try to use only the best cs in the field. Therefore we reject all stars that could not be measured on every image, faint stars with low S/N and variable stars which could introduce disturbing signals to the data. With the remaining objects we calculated the artificial cs. Finally this artificial cs is compared to TrES-2 to get the differential magnitudes for every image.\\ Then we correct for systematic effects by using the Sys-Rem detrending algorithm which is proposed by \cite[Tamuz et al. (2005)]{Tamuz_etal05} and implemented by Johannes Koppenhoefer. Figure \ref{TrES_Sysrem} shows the same transit event of TrES-2 before and after using Sys-Rem.
\\ To determine the time of the centre of the transit of TrES-2 we approximate a theoretical light curve on the observed light curve (see the right panel in Figure \ref{TrES_Sysrem}). To get the best fit we compare the theoretical light curve with the observed light curve until the $\chi^{2}$ is minimal.

\section{Transit timing residuals}

\begin{figure}[h]
  \centering
  \includegraphics[width=1\textwidth]{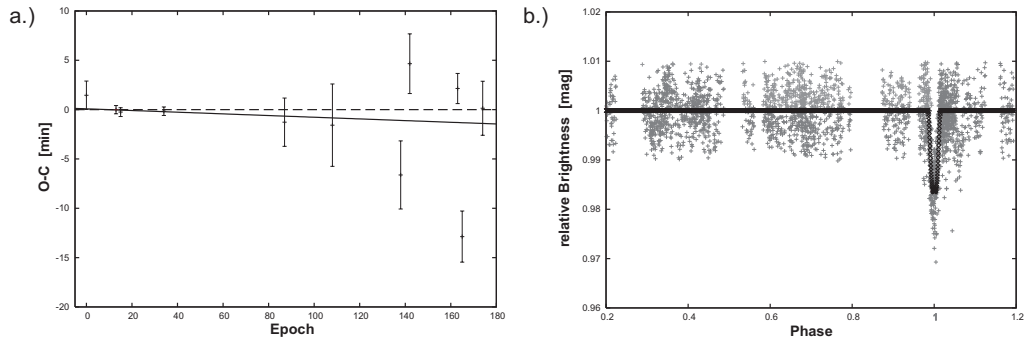}
  \caption{a.) Transit timing residuals for TrES-2. The dashed line shows the ephemeris given by \cite[Holman et al. (2005)]{Holman_etal07}. The best-fitting line (solid line) is plotted, representing the updated ephemeris given in equation \ref{Elemente:TrES2}. --- b.) More than 3000 individual observations of TrES-2 from March to October 2007 calculated in one phase according to the updated ephemeris.}
  \label{O_C_TrES2}
\end{figure}

In addition to the observation of TrES-2 from Gro{\ss}schwabhausen we found four transit times in the literature. These altogether 11 transits are summarized in the Table \ref{transit_times}.
\begin{table}
\centering
\caption{Summary of all known transit times of TrES-2.}
\label{transit_times}
\begin{tabular}{|c|c|c|c|}
\hline
Observer & Date & Time of Midtransit [HJD] & error  [d] \\ \hline
TrES-Network$^{1}$ & 10.08.2006 & 2453957.63580 & 0.00100 \\ 
\textit{TLC-Project}$^{2}$ & 11.09.2006 & 2453989.75286 & 0.00029 \\
& 16.09.2006 & 2453994.69393 & 0.00031 \\
& 02.11.2006 & 2454041.63579 & 0.00030 \\ 
Gsh & 13.03.2007& 2454172.57793 & 0.00170 \\
& 03.05.2007 & 2454224.46077 & 0.00290 \\
& 17.07.2007 & 2454298.57589 & 0.00240 \\
& 26.07.2007 & 2454308.46621 & 0.00210 \\
& 16.09.2007 & 2454360.34750 & 0.00106 \\
& 21.09.2007 & 2454365.27832 & 0.00180 \\
& 14.10.2007 & 2454387.52294 & 0.00190 \\ \hline
\end{tabular}
\vspace*{0.3cm}
\\
\begin{footnotesize}
$^{1}$from \cite[O'Donovan et al. (2006)]{ODonovan_etal06}\\
$^{2}$from \cite[Holman et al. (2005)]{Holman_etal07}
\end{footnotesize}
\end{table}
\\ We used the ephemeris of \cite[Holman et al. (2005)]{Holman_etal07} 
$T_{\mathrm{c}}(E)=(2453957.63479\,+\,E\cdot 2.470621)\,\mathrm{d}$
to compute ''observed minus calculated'' (O-C) residuals for all 11 transit times. Figure \ref{O_C_TrES2} shows the differences between the observed and predicted times of midtransit, as a function of epoch. The dashed line represents the ephemeris given by \cite[Holman et al. (2005)]{Holman_etal07}.
We found a negative trend in this (O-C)-diagram. Thus, we refine the ephemeris. The resulting ephemeris which represent our measurements best is
\begin{equation}
\label{Elemente:TrES2} 
T_{\mathrm{c}} = T_{0}+P\cdot E
\end{equation}
with:
\begin{eqnarray}
T_{0} &=& (2453989.75286 \pm 0.00029)\,\mathrm{d} \nonumber\\
P &=& (2.470615 \pm 0.00002)\,\mathrm{d} \nonumber
\end{eqnarray}

\section{Discussion and Outlook}

During the observations of the transiting extrasolar planet TrES-2 at our university observatory in Gro{\ss}schwabhausen with the 25\,cm Cassegrain telescope equipped with the optical CCD camera CTK we obtained a timing accuracy of $\sim$2\,min. The timing residuals  are not consistent with zero within the measurement errors. The second last data point is 3$\sigma$ deviant. The deviations to both sides of zero could be a first indication of timing anomalies caused by additional planets or moons. We will continue observing TrES-2 to confirm these transit time variations. Therefore, we work on methods to improve the accuracy of our transit times.\\ This year we get a new CCD camera for the Schmidt focus of the 90\,cm reflector. This camera will have a smaller pixel scale and a higher sensitivity. Part of the preparation for the new camera is the improvement of the software for relative photometry. Our transit observations will benefit strongly from the new camera.\\ The transit observations with the AIU Jena telescope in Gro{\ss}schwabhausen provide anchors for future searches for transit time variations.

\end{document}